\documentclass[12pt]{extarticle}
\usepackage{mathptmx}
\usepackage{graphicx}
\usepackage{amssymb}
\usepackage{latexsym}
\usepackage{authblk}
\title{New Trend in Quantization Methods}
\author{Laure Gouba}
\affil{Abdus Salam International Centre for Theoretical Physics\\ ICTP - Strada Costiera, 11, I - 34151 Trieste Italy\\
Email: lgouba@ictp.it}
\date{\today}
\begin{document}
\maketitle

\section{Introduction}

This paper is a slightly reduced and concise version of a course given at Geonet2019 school on {\it New Trends in Mathematical Methods for Physics } at IMSP in Rep. of Benin, May 2019.
The notes are intended to provide the reader with an introduction of a more recent procedure of quantization that is a generalization of the coherent states quantization procedure, highlighting the link between symplectic geometry and classical mechanics.

The topic and the goal of the Geonet2019 school motivated the choice of the title of the course as {\it New Trend in Quantization Methods}. It is also a coincidence of a current research interest in integral quantization. 

Quantization is in general understood as a correspondence between a classical and a quantum theory and in this sense, one also talks about dequantization, which is a procedure by which one starts with a quantum theory and arrives back to its classical counterpart. Due to the close connection between the transition process from classical to quantum mechanics and the geometry of the configuration and phase spaces, we highlight symplectic geometry and its link with classical mechanics.

The early notion of quantization was based on the following simple technique for quantizing a classical system: let $q^i, \: p_i,\; i= 1,2\ldots n$, be the canonical position and momenta, respectively, for a classical system with $n$ degrees of freedom. 
Their quantized counterparts, $\hat q^i, \hat p_i\; i= 1,2\ldots n$ are to be realized as operators on the Hilbert space $\mathcal{H} = L^2(\mathbb{R}^n, d\bold x)$ by the prescription 
\begin{equation}
(\hat q^i\psi)(\bold x) = x^i\psi(\bold x);\quad 
(\hat p_i\psi)(\bold x) = -i\hbar\frac{\partial}{\partial x^i}\psi(\bold x), 
\quad \psi  \in \mathcal{H}.
\end{equation} 
This simple procedure is known as canonical quantization.
 
Classical mechanics is formulated on phase space. 
In general, we may call a symplectic space $(\Gamma, \Omega)$, equipped with a Poisson bracket emanating from $\Omega$ a classical mechanics. Details about the definition of symplectic space is given in section \ref{sec1}. Let's consider $\mathcal{A}(\Gamma)$ a suitably large subalgebra of $C^\infty(\Gamma)$, equipped with the product $f\circ_{cl}g = \{f,g\}$, where 
$C^\infty$ denotes the space of infinitely differentiable functions. Let $f\in \mathcal{A}(\Gamma)$ and let $\hat f$ be its quantized counterpart. Then, according to the general rules of quantization (due to Dirac), we should require that the quantization map $f\mapsto \hat f$ satisfies the properties:
\begin{itemize}
\item $f\mapsto \hat f$ be real linear. 
\item For some appropriate subset $\{f\}\subset \mathcal{A}(\Gamma)$, which is closed under the Poisson bracket operation and chosen for physically motivated reasons, the corresponding quantized operators act irreducibly on $\mathcal{H}$;
\item 
$I_d \mapsto \mathbb{I}_d$, where $\forall x\in\Gamma,\: \mathbb{I}_d(x) = 1$ and $\mathbb{I}_d$ is the identity operator on $\mathcal{H}$; 
\item for a chosen set of functions $f,g\ldots,$
\begin{equation}
\{f,\;g\}\longrightarrow \frac{1}{i\hbar}[\hat f, \hat g],
\end{equation}
where $[\hat f,\hat g]$ is the commutator of the operators $\hat f$ and $\hat g$ on $\mathcal{H}$.
\end{itemize}

In the past five decades, a large number of techniques have been 
elaborated to make this transition from classical mechanics to quantum mechanics \cite{ali1, ali2}. The point of departure is always an analysis of the geometrical structure of either the classical phase space or the classical configuration space. There is, however no general theory of quantization presently available which is applicable in all cases, and indeed, often the technique used to quantize has to be tailored to the problem in question. 

We start with an overview of symplectic geometry in section \ref{sec1} and its link with classical mechanics in section \ref{sec2}. In section \ref{sec3}, we introduce the method of quantization  generically called integral quantization. Section \ref{sec4} is about existing quantization methods in the literature and why the integral quantization is seen as a new trend in  quantization methods.

\section{Overview of symplectic geometry}\label{sec1}

\subsection{About geometry}

{\it Geometry} is a branch of mathematics that deals with the measurement, the questions of shape, size, lines, angles surfaces and solids, the relative position of figures and the properties of space. {\it Euclidean geometry} is geometry in its classical sense. The mandatory educational curriculum of the majority of notions includes the study of points, lines, planes, angles, triangles, solid figures, circles and analytic geometry ( coordinate geometry or cartesian geometry ). {\it Differential geometry} is geometry studied via advanced calculus. It applies not just on euclidean space but on general smooth manifolds. It uses the techniques of differential calculus, integral calculus, linear algebra and multilinear algebra to study problems in geometry. {\it Riemannian geometry} is the oldest branch of differential geometry, and the branch that most directly generalizes Eulidean geometry: it is concern with lengths and angles. In mechanics, Riemannian geometry is used to describe kinetic energy; and also the paths followed by particles in the absence of external forces, which are geodesics.
{\it Symplectic geometry} is very different: it is not concerned with lengths or angles, but it does generalize area. It is an even dimensional geometry. It lies on even dimensional spaces and measures the sizes of 2-dimensional objects rather than 1-dimensional lengths and angles that are familiar from euclidean and riemannian geometry. Symplectic geometry has its origin in the Hamiltonian formulation of classical mechanics where the phase space of certain classical systems takes the structure of symplectic manifold. Manifolds are generalizations of the familiar ideas of lines, planes and their higher dimensional analogs; in other words a manifold is basically a smooth curve or surface or higher-dimensional analogs.

\subsection{Manifolds}

A real complex n-dimensional {\it manifold} M is a space which looks like an euclidean space $\mathbb{R}^n (\mathbb{C}^n)$ around each point,
but in which the global structure may be more complicated. The notion of dimension is important when discussing about manifolds.
Examples of one dimensional manifolds include a line, a circle, and two separate circles. Examples of two-dimensional manifolds include
a plane, a disk, the surface of a sphere, and the surface of a torus. Manifolds allow more complicated structures to be expressed and understood in terms of the relatively well-understood properties of simpler spaces.

More precisely, a manifold is defined by introducing a set of neighborhoods $U_i$ covering M, where each $U_i$ is a subspace of $\mathbb{R}^n(\mathbb{C}^n)$. A smooth manifold $M$ is a set of points together with a finite (or perhaps countable ) set of subsets $U_\alpha \subset M$ and one-to-one mappings $\phi_\alpha: U_\alpha \rightarrow \mathbb{R}^n$ such that 
\begin{enumerate}
\item $\cup_\alpha U_\alpha =  M $;
\item For every nonempty intersection $U_\alpha \cap U_\beta$ the set $\phi_\alpha(U_\alpha \cap U_\beta)$ is an open subset of $\mathbb{R}^n$ and the one -to- one mapping $\phi_\beta\circ \phi_\alpha^{-1}$ is a smooth function on $\phi_\alpha(U_\alpha \cap U_\beta)$.
\end{enumerate}

For example the boundary of a line segment is two end points and the boundary of a disk is a circle. Thus we may in general, determine another manifold of dimension $(n-1)$ by taking the boundary of an $n$ -manifold. The boundary of $M$ is denoted $\partial M$. The boundary of a boundary is always empty $\partial \partial M  = \emptyset$. 

One of the most important concepts used to study the properties of a manifold M is the {\it tangent space} $T_x(M)$ at a point $x\in M$. The tangent space $T_x(M)$ is thus defined as the vector space spanned by the tangents at $x$ to all curves passing through $x$ in the manifold. No matter how curved the manifold may be, $T_x(M)$ is always an n-dimensional vector space at each point $x$. A basis of the vector space $T_x M$ may be obtained by using the gradient operator, written as $\nabla = (\partial/\partial x^1, \partial/\partial x^2, \ldots, \partial /\partial x^n)$ in local coordinates. The disjoint union of tangent spaces to M at the points $q\in M$ given by 

\begin{equation}
TM =  \cup_{q\in M}T_q( M),
\end{equation}
is a vector space called the {\it tangent bundle} to $M$ and is denoted as $TM$. 

A {\it vector field} $X$ on a manifold $M$ is a map: $M \rightarrow TM$ that assigns a vector $X(x)$  at the point $x \in M$. The real vector space of vector fields on M is denoted $\mathfrak{X}(M)$. In the local coordinates a vector field $X$ has components $X^j$ given by 
\begin{equation}
X = X^j\frac{\partial}{\partial x^j} \equiv X^j\partial_j;\quad j = 1,2,\ldots n.
\end{equation}

Let $f: M \rightarrow \mathbb{R} $ be a smooth, real-valued function on an n-dimensional manifold $M$. The {\it differential} of $f$ at a point $x\in M$ is a linear map $df(x): T_x M\mapsto \mathbb{R}$. Relative to the local coordinate basis of the tangent space $T_x M$, one may write the dual basis as $dx^k, \: k = 1,2,\ldots , n,$ so that, in familiar notation, the differential of a function $f$ is given by 
\begin{equation}
df = \frac{\partial f}{\partial x^k}dx^k.
\end{equation}

Being a linear map from the tangent space $T_x M$ to $\mathbb{R}$ the differential defines the space $T^\star_x M$ dual to $T_x M$. The dual space $T^\star_x M$ is called the {\it cotangent space} of $M$ at $x$. The elements of $T^\star_x M$ are called {\it one-forms}. A one-form field $A$ on $M$ is a one-form $A(x) \in T^\star_x M $ for each value of $x\in M$. The union of cotangent spaces $T^\star_x M$ over all $x\in M$ is the {\it cotangent bundle}, denoted $T^\star M$. The dual of the tangent bundle $TM$ 
is the cotangent bundle $T^\star M$.

\subsection{Differential forms and exterior calculus}

Differential forms of higher degree may be constructed locally from one-form basis $dx^j, j= 1,2,\ldots n,$ by composition with the {\it wedge product}, or exterior product, denoted by the symbol $\wedge$.
 The geometric construction of higher-degree forms is intuitive and the wedge product is natural, if one imagines first composing the one-form basis as a set of line elements in space to construct oriented surface elements as two forms $dx^j\wedge dx^k$, then volume elements as three forms $dx^j\wedge dx^k\wedge dx^l$.
For these surface and volume elements to be oriented, the wedge product must be antisymmetric. That is, $dx^j\wedge dx^k = -dx^k\wedge dx^j$ under exchange of the order in a wedge product.
A 2-form is an expression build using wedge products of pairs in 1-forms.
The properties of the wedge product among differential forms in n dimensions are:
\begin{itemize}
\item $\alpha \wedge \beta$ is associative: $\alpha\wedge (\beta\wedge\gamma) = (\alpha\wedge\beta)\wedge \gamma$;
\item $\alpha\wedge \beta $ is bilinear in $\alpha$ and $\beta$;
\item $\alpha\wedge \beta$ is anticommutative: $\alpha\wedge \beta = (-1)^{kl}\beta \wedge\alpha$, where $\alpha$ is a k-form and $\beta$ is an l-form. The prefactor $(-1)^{kl}$ counts the signature of the switches in sign required in reordering the wedge product so that its basis indices are strickly icreasing, that is they satisfy $i_1<i_2<\ldots < i_{k+l}$.
\end{itemize}

In exterior calculus, the operation of {\it contraction} denoted as 
$\lrcorner$ introduces a pairing between vector fields and differential forms. Contraction is also called substitution of a vector field into a differential form. The rule for contraction of a vector field with a differential form develops from the relation 
\begin{equation}
\partial_j\lrcorner\; dx^k = \delta_j^k,
\end{equation}
in the coordinate basis $e_j = \partial_j :=\partial/\partial x^j$ and its dual basis $e^k =  dx^k$. 

The contraction of a vector field with a one-form yields the dot product, or inner product and th contraction between a covariant vector and a contravariant vector is given by 
\begin{equation}
X^j\partial_j\lrcorner\; v_kdx^k = v_k\delta_j^k X^j = v_j X^j.
\end{equation}
Let $\alpha$ be a k-form and $\beta$ be a one-form on a manifold $M$ and let $X = X^j\partial_j$ be a vector field. Then the contraction of $X$ through the wedge product $\alpha\wedge\beta$ satisfies 
\begin{equation}
X\lrcorner\; (\alpha\wedge\beta) = (X\lrcorner\;\alpha)\wedge \beta + (-1)^k\alpha\wedge ( X\lrcorner\;\beta).
\end{equation}
Beside the hook notation with $\lrcorner$, one also finds in the literature the following two alternative notations for the contraction of a vector field $X$ with $k$-form $\alpha \in \Lambda^k$ on a manifold $M$:
\begin{equation}
 X\lrcorner\;\alpha = i_X\alpha =  \alpha (X, \underbrace{\cdot,\cdot, \ldots, \cdot}) \in \Lambda^{k-1}.
\end{equation}
In the last alternative, one leaves a dot $(\cdot)$ in each remaining slot of the form that results after contraction. For example $X_H\lrcorner\;\Omega = \Omega (X_H,\cdot) = -\Omega(\cdot, X_H)$. 

\subsection{Symplectic vector space and symplectic manifold}

A symplectic vector space $(V, \Omega)$ is a real vector space $V$, together with an antisymmetric, non degenerate bilinear form 
$\Omega$ on $V$. This means that $\Omega: V\times V \rightarrow \mathbb{R}$ is a bilinear map, satisfying 
\begin{itemize}
\item $\Omega(a,b) = - \Omega (b,a)$, $\forall a, b\in V$; 
\item for fixed $a\in V$, $\Omega (a,b)= 0$, for all $a\in V$, implies that $a =0$.
\end{itemize}
Let $\textrm{dim} (V) = N$, and suppose that $\{e_i\}_{i=1}^N$ is a basis for $V$. Denotes by $V^\star$ the dual space of $V$. Then $\textrm{dim}(V^\star) = N$ and denote by $\{{\hat e}^i\}_{i=1}^N$ the dual basis of $V^\star$, we have $\langle {\hat e}^i; e_j\rangle = \delta_j^i,\: i,j = 1,2,\ldots, N,$ 
$\Omega$ can be written as 
\begin{equation}
\Omega = \frac{1}{2}\sum_{i,j =1}^N \omega_{ij}\hat e^i\wedge \hat e^j = \sum_{i,j =1}^N\omega_{ij}\hat e^i\otimes \hat e^j,
\end{equation}
where $\omega_{ij}$ are the elements of an $N\times N$ real, antisymmetric matrix $\bf \omega$ which is nonsingular, ${\bf \omega} = -{\bf \omega}^\intercal$, det ${\bf \omega} \neq 0$.
For $a,b\in V$, such that $\displaystyle{a =\sum_{i=1}^Na^ie_i,\;b= \sum_{i=1}^Nb^ie_i}$, we get 
\begin{equation}
\Omega (a,b):=b\lrcorner(a\lrcorner\;\Omega) = \sum_{i,j =1}^Na^i\omega_{ij}b^j.
\end{equation}
It can be shown that every symplectic vector space $(V, \Omega)$
is necessarily of even dimension, means $N = 2n$, and that there exists a basis $\{u_i,v_i\}_{i=1}^n$ in V for which 
\begin{equation}\label{eq1}
\Omega (u_i,v_j)= \delta_{ij}, \quad
\Omega (u_i,u_j) = \Omega(v_i,v_j)= 0, i,j = 1,2,\ldots n.
\end{equation} Such a basis is called symplectic frame. 

Let $\{\hat u^i,\hat v^i\}_{i=1}^n$ be the dual basis in $V^\star$, 
chosen so that 
\begin{equation}
\{\hat u^i; u_j\} = \{\hat v^i; v_j\}=\delta_j^i, \:
\{\hat u^i; v_j\} = \{\hat v^i; u_j\} = 0\; i,j =1,2,\ldots n.
\end{equation}
 Symplectic spaces are easily constructed, starting with any real, $2n$-dimensional vector space $V$ and an antisymmetric, non-singular $2n\times 2n$ matrix $\bf \omega$.
 
Let $(V,\Omega)$ be a symplectic vector space and $F\subset V$ a subspace. The symplectic complement of $F$ is the subspace 
\begin{equation}
F^\perp = \{ a \in V\ \backslash \Omega(a,b) = 0, \forall b \in F \}
\end{equation}
If $F$ and $G$ are two subspaces of $V$, then the following results are easy to prove:
\begin{itemize}
 \item $F\subset G \Rightarrow  G^\perp \subset F^\perp$;
 \item $(F^\perp)^\perp = F$;
 \item $(F+G)^\perp = G^\perp \cap G^\perp$;
\item $(F\cap G)^\perp = F^\perp + G^\perp$;
\item $\textrm{dim} (V) = \textrm{dim} (F) + 
\textrm{dim }(F^\perp)$.
\end{itemize}
A subspace $F$ of a symplectic vector space $(V,\Omega)$ is said to be 
\begin{itemize}
\item isotropic $\Longleftrightarrow  F\subseteq F^\perp $;
\item coisotropic $\Longleftrightarrow  F^\perp \subseteq F $ ;
\item symplectic $\Longleftrightarrow  F\cap  F^\perp = \{0\} $; \item Lagrangian $\Longleftrightarrow  F = F^\perp $.
\end{itemize}
From equation (\ref{eq1}) it follows that the subspaces $Q$ and $Q'$, generated by $\{v_i\}_{i=1}^n$ and $\{u_i\}_{i=1}^n$, respectively, are both Lagrangian. Hence Lagrangian subspaces always exist. {A linear canonical transformation of a symplectic vector space $(V, \Omega)$ is a linear map $S: V \rightarrow V$, such that 
\begin{equation}
\Omega (S a, S b) = \Omega (a,b),
\end{equation}
for all $a,b \in V$. The group of all such transformations is denoted $SP(V, \Omega)$}.

A symplectic manifold is a pair $(M, \omega)$ where
$M$ is a smooth real $m$-dimensional manifold without boundary
$\omega$ is a closed non-degenerate two-form. The symplectic form $\omega$ is {\it closed} means $d\omega = 0$, where
$d$ is the exterior differential defined by 
\begin{equation}
d: \Omega^k(M)\rightarrow \Omega^{k+1}(M),\quad d^2 = 0,
\end{equation} 
on differential forms on $M$. The {\it non-degeneracy} means that at each point $x\in M$ the antisymmetric matrix $\omega_x$ is non-degenerate, means $det(\omega_x)\neq 0,\; \forall x\in M$.

The most important example of a symplectic manifold is a cotangent bundle $M = T^\star Q$, that is nothing but the traditional phase space of classical mechanics, $Q$ being the configuration space in that context.
A cotangent bundle has a canonical symplectic two-form which is globally exact, $\omega = d\theta$. Any local coordinate system $\{q^k\}$ on $Q$ can be extended to a 
coordinate system $\{q^k, p_k\}$ on $T^\star Q$ such that $\theta$ 
and $\omega$ are locally given by 
\begin{equation}\label{cotg}
\theta = p_kdq^k,\quad \omega = dp_k\wedge dq^k.
\end{equation}

The non-degeneracy condition of $\omega$ has several important consequences:
it implies that $M$ is even-dimensional, $m =2n$; as $\omega$ is invertible, at each point $x\in M$ it gives an isomorphism between the tangent and cotangent spaces of $M$ at $x$, 
$\omega_x: T_x M \sim T^\star_x M$, expressed in local coordinates 
as $X^i \mapsto X^i\omega_{ik}$; the existence of $\omega$ allows us to associate a vector field $X_f$ to every function $f\in C^\infty (M)$ via $X_f\lrcorner\; \omega = -df$.
The vector field $X_f$ is the symplectic gradient of $f$, and is known as the Hamiltonian vector field of $f$. 
$X_f$ generates a flow on $M$  which leaves $\omega$ invariant as the Lie derivative of $\omega$ along $X_f$ is zero, 
\begin{equation}
L(X_f)\omega \equiv dX_f\lrcorner\; \omega + X_f\lrcorner\; d\omega = -ddf =0 .
\end{equation}
Via $X_f\lrcorner\; \omega = -df$, the symplectic form provides an antisymmetric pairing $\{f,g\}$ between functions $f,g$ on M called the Poisson bracket of $f$ and $g$. It is defined by 
\begin{equation}
\{f,g\}:= \omega (X_f,X_g)\in C^\infty (M),
\end{equation}
and describes the change of $g$ along $X_f$ (or vice versa), 
\begin{equation}
\{f,g\} = X_g\lrcorner\; X_f\lrcorner\; \omega = X_f\lrcorner\; dg= L(X_f)g.
\end{equation}
The Poisson bracket satisfies the Jacobi identity 
\begin{equation}
\{f,\{g,h\}\} =  \{\{f,g\},h\} + \{g,\{f,h\}\},
\end{equation}
$(C^\infty(M), \{.,.\})$ has the structure of an infinite dimensional Lie algebra.
An identity which relates the Lie algebras of vector fields and functions on $M$ is 
\begin{equation}
[X_f, X_g] = X_{\{f,g\}}, 
\end{equation}
which shows that the Hamiltoninan vector fields also form an infinite dimensional Lie algebra. This identity is an illustration of the quantization paradigm regarding the map $f\to X_f$ as an assignment of differential operators to functions.

A subspace $(V, \omega\vert_V)$ of a symplectic vector space $(W,\omega)$ is called istropic if $\omega\vert_V = 0$. By linear algebra, an isotropic subspaces of $W$ has dimension at most 
$\frac{1}{2}\textrm{dim} (W)$, and in that case $V$ is called a Lagrangian subspace of $W$.

\section{ Symplectic geometry and classical mechanics}\label{sec2}

During the seventeenth and eighteenth Century, classical mechanics  was very geometrical. Although Newton invented the calculus in order to formulate and solve physical problems, many of his arguments made heavy use of euclidean geometry.
After Newton, there came a period of {\it m\'ecanique analytique}, during which Lagrange could illustrate his treatise on mechanics without pictures, the developement of analytic techniques for the explicit solution of the differential equations describing mechanical systems has been continued by Euler, Lagrange, Jacobi and Hamilton. The contributions of Poincar\'e and Birkhoff has given to geometry a new role in mechanics. Symplectic geometry is of interest today because of a series of remarkable "transforms" which connect it with various areas of analysis. Symplectic geometry is the adequate mathematical framework for describing the Hamiltonian version of classical mechanics. In that sense, it is also the most suitable starting point for a geometrization of the canonical quantization procedure. Classical mechanics is formulated on phase space, a $2n$-dimensional real vector space $\sim \mathbb{R}^{2n}$ with coordinates $q^1,\ldots q^n, p_1,\ldots p_n$ describing the position and the momentum (velocity) of the particles involved.
The dynamics (time evolution) of the system is governed by Hamiltonian's equations
\begin{equation}\label{eqm1}
\frac{d}{dt}q^k = \frac{\partial H}{\partial p_k};\quad
\frac{d}{dt}p_k = -\frac{\partial H}{\partial q^k},
\end{equation} 
with $H(q^k, p_k)$, is the Hamiltonian, a function on phase space describing the energy of the system.

If $H$ does not depend on time explicitly, the equations of motion in (\ref{eqm1}) imply that $H$ is conserved along any trajectory in phase space, 
\begin{equation}\label{eqm2}
\frac{d}{dt}H =\frac{\partial H}{\partial q^k}\dot{q}^k 
+ \frac{\partial H}{\partial p_k}\dot{p}_k 
= \frac{\partial H}{\partial q^k}\frac{\partial H}{\partial p_k}
- \frac{\partial H}{\partial q^k}\frac{\partial H}{\partial p_k} = 0.
\end{equation}
The evolution of any other function $f$ on phase space (observable)is given by 
\begin{equation}\label{eqm3}
\frac{d}{dt} f = \frac{\partial f}{\partial q^k}\frac{\partial H}{\partial p_k}
- \frac{\partial f}{\partial p_k}\frac{\partial H}{\partial q^k},
\end{equation}
any function $f$ on phase space is in involution with the Hamiltonian, means $\{H,f\} = 0$, a constant of motion.
The equations (\ref{eqm1}-\ref{eqm2}-\ref{eqm3}) characterise Hamiltonian mechanics and arise naturally if we think of 
$\mathbb{R}^{2n}$ as the cotangent bundle $T^\star \mathbb{R}^n$ of the configuration space $\mathbb{R}^n$, with the canonical symplectic form (\ref{cotg}). In that case the Hamiltonian vector field $X_f$ of a function $f(q^k, p_k)$ is 
\begin{equation}
X_f = \frac{\partial f}{\partial p_k}\frac{\partial}{\partial q^k}
- \frac{\partial f}{\partial q^k}\frac{\partial}{\partial p_k}.
\end{equation}
It is easy to verify that $X_f\lrcorner\; dp_k\wedge dq^k = -df$
Therefore the Poisson bracket is 
\begin{equation}
\{f,g\} = \frac{\partial f}{\partial p_k}\frac{\partial g}{\partial q^k}
- \frac{\partial f}{\partial q^k}\frac{\partial g}{\partial p_k}.
\end{equation}
The canonical Poisson brackets (classical canonical commutations relations) between the coordinates and momenta are
\begin{equation}
\{q^k,q^l\} = \{p_k,p_l\} = 0;\quad \{p_k,q^l\} = \delta_k^l.
\end{equation}
The function $q^k$ and $p_l$ form a complete set of observables.

The equations (\ref{eqm1}) - (\ref{eqm2}) - (\ref{eqm3}) can be written as 
\begin{itemize}
\item (\ref{eqm1}) $\longleftrightarrow \frac{d}{dt}q^k = \{H, q^k\},\quad \frac{d}{dt}p_k = \{H, p_k\}  $,
\item (\ref{eqm2}) $\longleftrightarrow \{H,H\} = 0$,
\item (\ref{eqm3}) $\longleftrightarrow \frac{d}{dt} f = \{H,f\} =  X_H f$ so that time evolution in classical mechanics is determined by the Hamiltonian vector field $X_H$ of the Hamiltonian H.
\end{itemize}
Very generally, we may call a symplectic space $(\Gamma, \Omega)$, 
equipped with a Poisson bracket emaning from $\Omega$ a classical mechanics. States in classical mechanics are probability measures $\mu$, defined on $\Gamma$, means Borel measures satisfying 
\begin{equation}
\int_\Gamma d\mu(x) = 1.
\end{equation}
 In a pure state, localised at $x= x_0$ is a $\delta$-measure concentrated at $x_0$: 
\begin{equation}
\delta_{x_0}(f) =  f(x_0), \: f\in C^\infty (\Gamma).
\end{equation}
Observables in classical mechanics are real valued functions 
$f:\Gamma \rightarrow \mathbb{R}$, and the value of an observable 
$f$, in a state $\mu$, is the expectation value:
\begin{equation}
\langle \mu; f\rangle = \int_\Gamma f(x)d\mu(x).
\end{equation}
Quantum mechanics, on the other hand, is based on a (complex and separable) Hilbert space $\mathcal{H}$. 

A pure state can be represented by a one-dimensional projection operator, 
\begin{equation}
\Phi = \frac{\vert \phi\rangle\langle\phi\vert}{||\phi||^2}, \phi\neq 0
\end{equation}
on $\mathcal{H}$. The space of all such projection operators, 
$\mathbb{C}\mathbb{P}(\mathcal{H})$, also has the structure of a manifold (generally infinite dimensional). A mixed state in quantum mechanics can be represented by a density matrix
\begin{equation}
\rho = \sum_{n=1}^N\alpha_n\vert n\rangle\langle e_n\vert,\quad \alpha_n\ge 0, \sum_{n=1}^N\alpha_n = 1, \quad N = \textrm{dim}(\mathcal{H}),
\end{equation}
where $\{e_n\}_{n=1}^N$ is an orthonormal basis in $\mathcal{H}$.
The observables of quantum mechanics are self-adjoint operators 
$A$, on the Hilbert space $\mathcal{H}$ and the value of $A$ in the state $\Phi$ is given by 
\begin{equation}
\langle \Phi; A\rangle = \langle \hat \phi\vert A\vert \hat \phi\rangle_{\mathcal{H}} = \textrm{tr}[\Phi A],
\end{equation}
$\hat\phi$ is any vector in the range of the projection operator 
$\Phi$, and $\langle \cdot\vert\cdot\rangle_{\mathcal{H}}$ denotes the scalar product in $\mathcal{H}$. For the mixed state $\rho$, 
\begin{equation}
\langle \rho; A\rangle = \textrm{tr}[\rho A] = \sum_{n=1}^N\alpha_n\langle e_n\vert Ae_n\rangle.
\end{equation}

Let's consider the set of classical observables to be a (non-associative) algebra with respect to the product
$f \circ_{cl} g = \{f,g\}$, given by the Poisson bracket and
the set of quantum observables to be (non-associative) algebra with respect to the commutator algebra
$A \circ_{qu} B = \frac{1}{i\hbar}[A,B]$. Let $\mathcal{A}(\Gamma)$ a suitably large subalgebra of $C^\infty(\Gamma )$ equipped with the product $f\circ_{cl} g$ and $\mathcal{S}(\mathcal{H})$ some suitably large class of self-adjoint operators on $\mathcal{H}$, equipped with the product $A \circ_{qu} B $.
{\it A quantization of the classical mechanics $(\Gamma,\Omega)$ 
would seem to imply finding an appropriate Hilbert space $\mathcal{H}$ and an algebraic homomorphism between $\mathcal{A}(\Gamma)$ and $\mathcal{S}(\mathcal{H})$}.

\section{Introduction to integral quantization}\label{sec3}
\subsection{Dirac ket notations in Euclidean plane}
Let's consider an orthonormal basis (or frame) of the Euclidean plane $\mathbb{R}^2$ defined by the two vectors (in Dirac ket notations)
\begin{equation}
\vec{i} \equiv \vert 0\rangle \equiv 
\left(\begin{array}{cc}
1\\
0
\end{array}
\right)\quad \textrm{and}\quad 
\vec{j} \equiv \vert \frac{\pi}{2}\rangle \equiv 
\left(\begin{array}{cc}
0\\
1
\end{array}
\right)
\end{equation}
and $\vert \theta\rangle$ denotes the unit vector with polar angle $\theta \in [0,\;2\pi)$.
This frame is such that 
\begin{equation}
\langle 0\vert 0 \rangle = 1 = 
\langle \frac{\pi}{2}\vert \frac{\pi}{2} \rangle \quad 
 \textrm{and}\quad  \langle 0\vert \frac{\pi}{2}\rangle= 0.
\end{equation}
The sum of their corresponding orthogonal projectors resolves the unit 
\begin{equation}
\mathbb{I} = \vert 0\rangle\langle 0\vert + \vert \frac{\pi}{2}\rangle\langle \frac{\pi}{2}\vert,
\end{equation}
that is equivalent to 
\begin{equation}
\left(
\begin{array}{cc}
 1 & 0\\ 0 & 1
\end{array}\right)  =
\left(
\begin{array}{cc}
 1 & 0\\ 0 & 0
\end{array}\right) +
\left(
\begin{array}{cc}
 0 & 0\\ 0 & 1
\end{array}\right).
\end{equation}
To the unit vector $\vert \theta\rangle = \cos\theta\vert 0\rangle + 
\sin\theta\vert\frac{\pi}{2}\rangle$ corresponds the orthogonal projector 
$P_\theta$ given by $P_\theta  = \vert \theta\rangle\langle \theta\vert$  and 
\begin{equation}
\vert \theta \rangle \langle \theta\vert = 
\left(\begin{array}{cc}
\cos\theta\\\sin\theta\end{array}\right)
\left( \cos\theta\:\: \sin\theta\right)
= \left(
\begin{array}{cc}
\cos^2\theta & \cos\theta\sin\theta\\
\cos\theta\sin\theta & \sin^2\theta
\end{array}\right)
\end{equation}
$P_\theta = \mathcal{R}(\theta)\vert \theta\rangle\langle \theta\vert\mathcal{R}(-\theta)$, where 
\begin{equation}
\mathcal{R}(\theta) = \left(
\begin{array}{cc}
\cos\theta  & -\sin\theta\\
\sin\theta & \cos\theta
\end{array}
\right)\in \textrm{SO}(2).
\end{equation}
\subsection{Sea star algebra}
An original approach to quantization based on operator valued measures is named generically integral quantization. 
The approach generalizes the coherent states quantization.
The five-fold symmetry of sea star (starfish) motivated the developpment in a comprehensive way the quantization of functions on a set with five elements which yields in particular the notion of  a quantum angle.
\begin{figure}[h]
\centering
 \includegraphics[width=6cm]{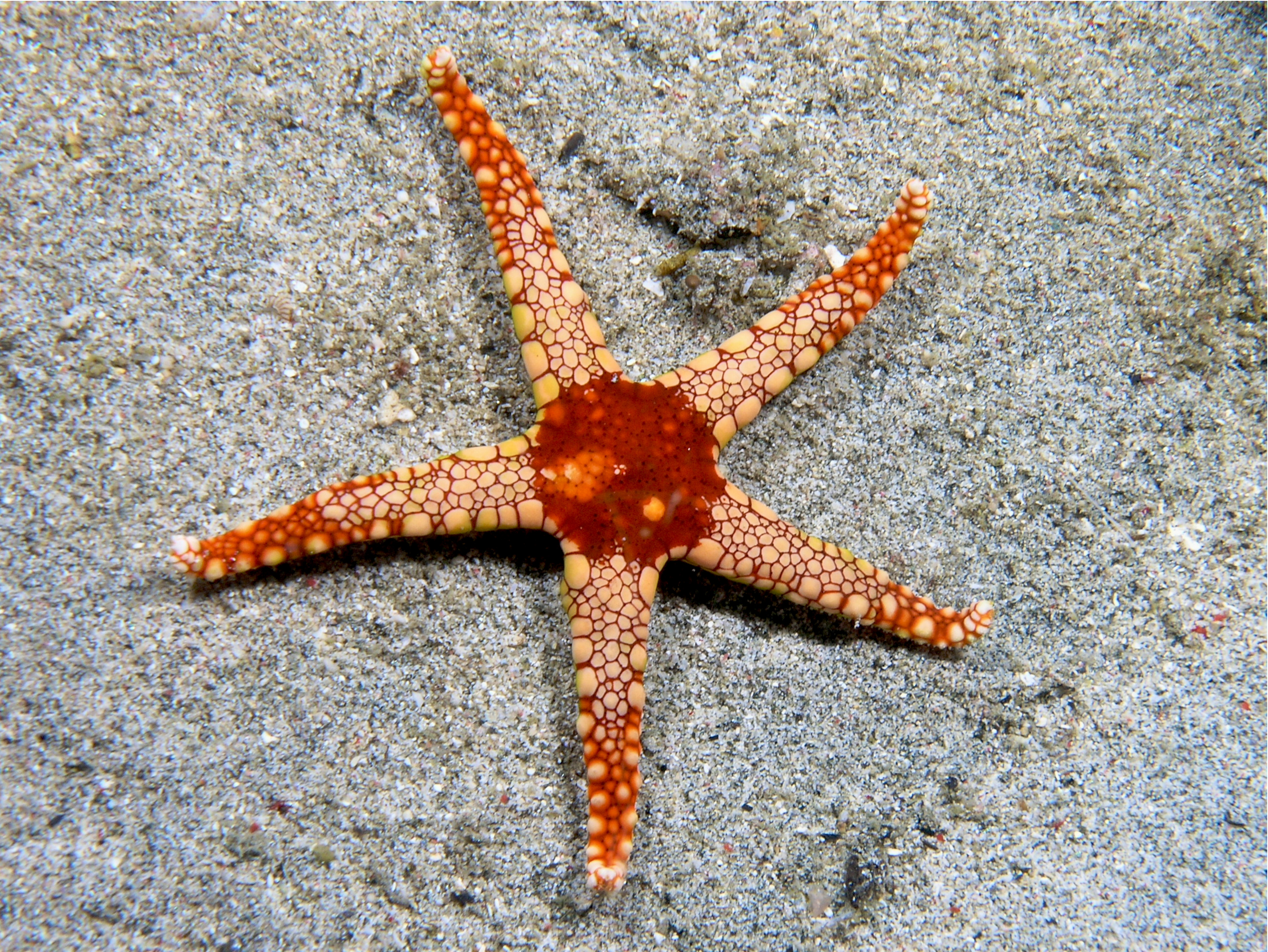}
\caption{A sea star as a five-fold frame for the plane}  \label{fig:seastar}
\end{figure}

 Sea stars are characterized by five arms, or rays, connected to a small round body. Sea stars detect light with five purple eyespots at the end of each arm. Sea stars typically show pentameral symmetry. The purpose of this introductory model is to enhance our intuition that quantization can be viewed as the analysis of a set from the point of view of a family of coherent states.

The pentagonal set of unit vectors (the arms) form a five-fold frame in the plane. The unit vectors (arms) are determined by $\vert \frac{2n\pi}{5}\rangle = 
\mathcal{R}(\frac{2n\pi}{5})\vert 0\rangle \equiv$ {\it coherent state}, $n = 0,1,2,3,4$ mod 5.
 We get the resolution of the identity:
\begin{equation}
\frac{2}{5}\sum_{n=0}^4\vert \frac{2\pi n}{5}\rangle\langle 
\frac{2\pi n}{5}\vert =  \left(\begin{array}{cc} 1 & 0\\
0 & 1\end{array}\right)\equiv \mathbb{I}.
\end{equation}
The property holds for any regular N-fold polygon in the plane
\begin{equation}
\frac{2}{N}\sum_{n=0}^{N-1}\vert \frac{2\pi n}{N}\rangle\langle 
\frac{2\pi n}{N}\vert =  \left(\begin{array}{cc} 1 & 0\\
0 & 1\end{array}\right).
\end{equation}
Let's make more precise the set $X$ explored by the sea star. 
Let's determine a probabilistic construction of the frame by using the Hilbert space structure of the plane. 

The sea star senses its possible orientations via the five angles $\frac{2\pi n}{5}$ and $X = \{0,1,2,3,4\}$ the set of orientations. 
It is equipped with discrete measure with uniform weight 
\begin{equation}
\int_x f(x)d\mu(x) = \frac{2}{5}\sum_{n=0}^4 f(n)
\end{equation}
Let us choose two orthonormal elements, $\phi_0(n) =\cos({2\pi n/5})$ and $\phi_1(n) = \sin({2\pi n/5})$ in the Hilbert space $L^2(X,\mu)$, and build the five unit vectors in the real two-dimensional Hilbert space, that is the Euclidean plane $\mathbb{R}^2  = \mathcal{H}$, with the usual orthonormal basis $\vert 0\rangle,\; \vert \frac{\pi}{2}\rangle$:
\begin{equation}
n\in X,\quad n\mapsto \vert n\rangle \equiv \vert \frac{2\pi n}{5} \rangle 
= \phi_0(n)\vert 0\rangle +\phi_1(n)\vert \frac{\pi}{2}\rangle.
\end{equation}
The operator $P_n = \vert\frac{2\pi n}{5}\rangle \langle \frac{2\pi n}{5}\vert $ acts on $\mathcal{H} = \mathbb{R}^2$
Given a $n_0 \in \{0,1,2,3,4\}$, one derives the probability distribution on $X \in \{0,1,2,3,4\}$ 
\begin{equation}
\textrm{tr}(P_{n_0} P_n) = \vert \langle \frac{2\pi n_0}{5}\vert \frac{2\pi n}{5}\rangle\vert^2 = \cos^2 \left(\frac{2\pi(n_0 -n)}{5}\right).
\end{equation}
The five unit vectors $\vert n\rangle\equiv \vert \frac{2\pi n}{5}\rangle$ resolve the identity in $\mathcal{H}$. They form a finite unit frame for analysing the complex-valued functions $n\mapsto f(n)$ on X through what we call a coherent state (CS) quantization:
\begin{equation}
f(n)\mapsto \int_X\vert x\rangle \langle x\vert f(x) d\mu(x) = \frac{2}{5}\sum_{n=0}^4 f(n)\vert\frac{2n\pi}{5}\rangle\langle \frac{2n\pi}{5}\vert \equiv A_f.
\end{equation}
 If we had chosen instead as a finite frame the orthonormal basis $\vert 0\rangle, \vert \pi/2 \rangle$, in $\mathbb{R}^2$, we would have obtained the trivial commutative quantization: 
\begin{equation}
(f(0), f(1))\mapsto A_f= \textrm{diag}(f(0), f(1)).
\end{equation}
This map should be analysed more precisely through spectral values of the $2\times 2$ symmetric matrix $A_f$, and also through CS mean values of $A_f$: 
\begin{equation}
n\mapsto \check{f}(n) := \langle \frac{2n\pi}{5}\vert A_f\vert \frac{2 n\pi}{5}\rangle  = \frac{2}{5}\sum_{m=0}^4 f(m)\cos^2\frac{2(n-m)\pi}{5}.
\end{equation}
More details can be read in reference \cite{gaz0}.

\subsection{Integral Quantization}

To be more mathematically precise and still remaining at an elementary or minimal level, quantization is a linear map $\mathcal{Q}:\mathcal{C}(X)\rightarrow \mathcal{A}(\mathcal{H})$ from a vector space $\mathcal{C}(X)$ of complex valued function $f(x)$ on a set $X$ to a vector space $\mathcal{A}(\mathcal{H})$ of linear operator $\mathcal{Q}(f)\equiv A_f$ in some complex Hilbert space $\mathcal{H}$, such that 
\begin{enumerate}  
\item to the function $ f = 1$ there corresponds the identity operator $\mathbb{I}$ on $\mathcal{H}$
\item to a real function $f\in \mathcal{C}(X)$ there corresponds 
a  (an essentially) self adjoint operator $A_f$ in $\mathcal{H}$.
\end{enumerate}

Let $(X,\nu) $ be a measure space.
 Let $\mathcal{H}$ be a complex Hilbert space and $x\in X,\; x\mapsto M(x)\in \mathcal{L}(\mathcal{H})$ an X-labelled family of bounded operators on $\mathcal{H}$ resolving the identity:
\begin{equation}
\int_X M(x) d\nu(x) = \mathbb{I},
\end{equation}
provided that the equality is valid in a weak sense, which implies $\nu$ integrability for the family $M(x)$.
 If the operators $M(x)$ are positive and have unit trace, they will be preferentially denoted by $M(x) = \rho(x)$ in order to comply with the usual notation for a density matrix in quantum mechanics.
The corresponding quantization of complex-valued functions $f(x)$ on $X$ is then defined by the linear map:
\begin{equation}
 f \mapsto A_f =  \int_X M(x)f(x)d\nu (x).
\end{equation}
This operator-valued integral is again understood in the weak sense, as the sesquilinear form, 
\begin{equation}
B_f(\psi_1,\psi_2) = \int_X \langle \psi_1\vert M(x)\vert\psi_2\rangle f(x)d\nu(x).
\end{equation}
The form $B_f$ is assumed to be defined on a dense subspace of $\mathcal{H}$.
 If $f$ is real and at least semi-bounded, the Friedrichs extension of $B_f$ univocally defines a self-adjoint operator.
If $f$ is not semi-bounded, there is no natural choice of a self-adjoint operator associated with $B_f$, in this case, we can consider directly the symmetric operator $A_f$, enabling us to obtain a self-adjoint extension (unique for particular operators)
 The question of what is the class of operators that may be so represented is a subtle one.
 One of the interesting aspects of this integral quantization scheme is precisely to analyse the quantization issues, particularly the spectral properties of operators $A_f$. Another appealing aspects is the possibility to quantize constraints. Integral quantization with two basic examples can be read in reference \cite{gaz1}.
 
\section{Survey on Quantization techniques}\label{sec4}

A review titled {\it Quantization Methods: A Guide for Physicists and Analysts } by Syed Twareque Ali and 
Miroslav Engli$\check{s}$ is available in the literature \cite{ali2}. 
In this review the authors present an overview of some of the better known quantization techniques found in the current literature and used both by physicists and mathematicians:
\begin{itemize} 
\item canonical quantization and its generalization: the early notion of quantization, Segal quantization, Borel quantization; \item geometric quantization; 
\item deformation quantization; 
\item Berezin and Berezin-Toeplitz quantization on K\"ahler manifolds;
\item prime quantization; 
\item coherent state quantization; 
\item some other quantization methods enumerated in the review.
\end{itemize}
The reader interested in understanding the methods enumerated may 
read in \cite{ali1, ali2}. Recently the integral quantization has been mostly investigated by Jean-Pierre Gazeau and his collaborators\cite{gaz1,gaz2,gaz3, gaz4} and with respect to the existing methods in the literature enumerated, we consider it as the new trend of quantization.

\section*{Acknowledgements}
L. Gouba would like to gratefully thank the organizers of the Geonet2019 school for the invitation and the hospitality.

\end{document}